\documentclass[fleqn,usenatbib,useAMS]{mnras}

\usepackage[T1]{fontenc}
\usepackage{ae,aecompl}
\usepackage{newtxtext,newtxmath}

\usepackage{graphicx}	
\usepackage{amsmath}	
\usepackage{amssymb}	
\usepackage{subfigure}
\usepackage{xfrac}
\usepackage{soul}
\usepackage{comment}
\usepackage{color}
\usepackage{multirow}
\usepackage{mathtools}
\usepackage{multicol}        
\usepackage{bm}		

\title[Why do massive discs exist?]{Why do extremely massive disc galaxies exist today?}

\author[R. A. Jackson et al.]
{R. A. Jackson$^{1}$\thanks{E-mail: r.jackson9@herts.ac.uk},
G. Martin,$^{1,2,3}$
S. Kaviraj,$^{1}$
C. Laigle,$^{4,5}$
J. E. G. Devriendt,$^{5}$ \newauthor 
Y. Dubois, $^{4}$ and
C. Pichon $^{4,6,7}$
\\
$^{1}$Centre for Astrophysics Research, School of Physics, Astronomy and Mathematics, University of Hertfordshire, Hatfield, AL10 9AB, UK\\
$^{2}$Steward Observatory, University of Arizona, 933 N. Cherry Ave, Tucson, 85719, AZ, USA\\
$^{3}$Korea Astronomy and Space Science Institute, 776 Daedeokdae-ro, Yuseong-gu, Daejeon 34055, Korea\\
$^{4}$Institut d'Astrophysique de Paris, Sorbonne Universit\'es, UMPC Univ Paris 06 et CNRS, UMP 7095, 98 bis bd Arago, 75014 Paris, France\\
$^{5}$Dept of Physics, University of Oxford, Keble Road, Oxford OX1 3RH UK\\
$^6$School of Physics, Korea Institute for Advanced Study (KIAS), 85 Hoegiro, Dongdaemun-gu, Seoul, 02455, Republic of Korea\\
$^7$Institute for Astronomy, University of Edinburgh, Royal Observatory, Blackford Hill, Edinburgh, EH9 3HJ, United Kingdom\\
}

\begin{document}
\label{firstpage}
\pagerange{\pageref{firstpage}--\pageref{lastpage}}
\maketitle

\begin{abstract}
Galaxy merger histories correlate strongly with stellar mass, largely regardless of morphology. Thus, at fixed stellar mass, spheroids and discs share similar assembly histories, both in terms of the frequency of mergers and the distribution of their mass ratios. Since mergers are the principal drivers of disc-to-spheroid morphological transformation, and the most massive galaxies typically have the richest merger histories, it is surprising that discs exist at all at the highest stellar masses (e.g. beyond the knee of the mass function). Using Horizon-AGN, a cosmological hydro-dynamical simulation, we show that extremely massive (M$_*$ $>$ 10$^{11.4}$ M$_\odot$) discs are created via two channels. In the primary channel (accounting for ~70$\%$ of these systems and ~8$\%$ of massive galaxies) the most recent, significant merger (stellar mass ratio $>$ 1:10)  between a massive spheroid and a gas-rich satellite `spins up' the spheroid by creating a new rotational stellar component, leaving a massive disc as the remnant. In the secondary channel (accounting for ~30$\%$ of these systems and ~3$\%$ of massive galaxies), a system maintains a disc throughout its lifetime, due to an anomalously quiet merger history. Not unexpectedly, the fraction of massive discs is larger at higher redshift, due to the Universe being more gas-rich. The morphological mix of galaxies at the highest stellar masses is, therefore, a strong function of the gas fraction of the Universe. Finally, these massive discs have similar black-hole masses and accretion rates to massive spheroids, providing a natural explanation for why a minority of powerful AGN are surprisingly found in disc galaxies.
\end{abstract}



\section{Introduction}

In the standard hierarchical paradigm, the morphological mix of massive galaxies is predicted to change from rotationally-supported discs to dispersion-dominated spheroids over cosmic time \citep[e.g.][]{Butcher1984,Dressler1997,Conselice2014}. Observations generally support this picture. While discs dominate the high redshift Universe, the morphologies of massive galaxies in the nearby Universe are mainly spheroidal in nature \citep{Bernardi2003,Wuyts2011,Kaviraj2014a,Kaviraj2014b,Conselice2014,Buitrago2014,Shibuya2015}. This disc-to-spheroid transformation is thought to be primarily driven by galaxy mergers \citep{Toomre1977,Barnes1992,Bournaud2007,DiMatteo2007,Oser2010,Kaviraj2010,Kaviraj2011,Dubois2013,Dubois2016,Lofthouse2017,Welker2018,Martin2018b}. The gravitational torques generated by mergers (especially `major' mergers, which have nearly equal mass ratios) remove stars from ordered rotational orbits and put them on random trajectories, producing dispersion-dominated remnants \citep[e.g.][]{Springel2005,Hilz2013,Font2017,Martin2018b}.

The role of mergers is considered to be progressively more important for more massive galaxies. At the very highest stellar masses, i.e. beyond the knee of the mass function (M$_*$ $>$ 10$^{10.8}$ M$_\odot$, see e.g. \citet{Li2009,Kaviraj2017}), the consensus view is that mergers are essential for actually achieving the enormous stellar masses of such systems \cite[e.g.][]{Faber2007,McIntosh2008,Cattaneo2011}. Since mergers typically create dispersion-dominated stellar components, it is not surprising that massive galaxies are dominated by spheroidal systems.

Interestingly, however, both observations \citep[e.g.][]{Conselice2006, Ogle2016, Ogle2019} and theory \citep[e.g.][]{Martin2018b} indicate that, even at the highest stellar masses, a significant minority of systems (e.g. more than 10 per cent at M$_*$ $>$ 10$^{11.4}$ M$_\odot$) surprisingly host significant disc components. For example, in the SDSS \citep{Abazajian2009}, \citet{Ogle2016,Ogle2019} find that a subset of the most optically luminous (M$_* = 0.3-3.4\times10^{11}$M$_{\odot}$) galaxies have clear disc components. They speculate that these `super spirals' may have formed via gas-rich major mergers between two spiral galaxies, with the gas of the two merging galaxies combining to form large gas discs which then create the discy stellar components. 

Other recent work supports the finding that such massive discs are relatively gas-rich \citep{Li2019} and indicates that these systems can be found in a variety of environments, including clusters \citep{Bogd2018,Li20192}. These studies suggest that these galaxies can even be the brightest galaxies in their respective groups and clusters, residing at the barycentres of these structures. Indeed, if such galaxies live in such high-density environments and host AGN \citep{Ogle2016}, they could provide a natural explanation for the minority of powerful radio AGN that apppear (somewhat surprisingly) to be hosted by discy systems \citep[e.g.][]{Tadhunter2016}. 

In the $\Lambda$CDM model, galaxy merger histories are a strong function of stellar mass, largely regardless of the morphology of the galaxy in question. The merger histories of extremely massive discs and spheroids are, therefore, very similar, both in terms of the total number of mergers they experience and the distribution of their mass ratios \citep[e.g][]{Martin2018b}. Since mergers typically destroy discs and create spheroids, it is surprising that any discs exist at all in this extreme mass regime. It is likely, therefore, that some peculiarity in the characteristics of their merger histories causes these massive discs to either survive or for discy (i.e. rotationally-supported) components to be regenerated. This is plausible as gas-rich mergers can
regenerate discs, as the gas produces new, rotationally-supported stellar components during the course of the merger event \citep[e.g.][]{Springel2005,Robertson2006,Governato2009,Hopkins2009,Font2017,Martin2018b,Peschken2019}. 

Here, we explain the origin of extremely massive disc galaxies in the nearby Universe, by exploring how details of their merger histories create these surprising systems, using the Horizon-AGN cosmological hydrodynamical simulation. It is worth noting that a cosmological simulation, such as the one used here, is essential for this exercise. While idealised and/or zoom-in simulations of galaxy mergers have often been used to study the merger process \citep[e.g.][]{Barnes1988,Hernquist1992,Bois2011,Athanassoula2016,Sparre2016,Sparre2017}, the parameter space sampled by these experiments is typically small and not informed by a cosmological model (so that the effects of environment and gas accretion from the cosmic web cannot be fully tested). 

The plan for this paper is as follows. In Section \ref{sec:horizon}, we describe the Horizon-AGN simulation which underpins this study. In Section \ref{sec:disc formation}, we explain the different formation channels that lead to the formation of extremely massive discs and explore whether these massive discs can explain the discy hosts of powerful AGN (which are otherwise typically hosted by spheroidal galaxies). We summarise our findings in Section \ref{sec:summary}.


\section{The Horizon-AGN Simulation} 
\label{sec:horizon}

We use the Horizon-AGN cosmological hydrodynamical simulation \citep{Dubois2014}, which employs \textsc{ramses} \citep{2002A&A...385..337T}, an adaptive mesh refinement (AMR) hydrodynamics code. The simulation offers a 100 $h^{-1}$ comoving Mpc$^3$ volume and uses WMAP7 $\Lambda$CDM initial conditions \citep{2011ApJS..192...18K}. Horizon-AGN contains $1024^3$ dark matter particles on an initial 1024$^3$ cell gas grid, which is refined using a quasi Lagrangian criterion, when 8 times the initial total matter resolution is reached in a cell. This refinement continues down to 1 kpc in proper units (which therefore sets the minimum cell size and the spatial resolution of the simulation). The simulation has a dark-matter mass resolution of $8\times 10^7$ M$_{\odot}$, gas mass resolution of $\sim10^7$M$_{\odot}$ and stellar mass resolution of $4\times 10^6$ M$_{\odot}$. 

Horizon-AGN employs prescriptions for both stellar and AGN feedback. Stellar feedback includes momentum, mechanical energy and metals from Type Ia/Type II supernovae (SNe), with the Type Ia SNe implemented following \citet{1986A&A...154..279M}, assuming a binary fraction of 5\% \citep{2001ApJ...558..351M}. Feedback from Type II SNe and stellar winds is implemented using \textsc{starburst99} \citep{1999ApJS..123....3L,2010ApJS..189..309L}, via the Padova model \citep{2000A&AS..141..371G} with thermally pulsating asymptotic branch stars \citep{1993ApJ...413..641V}. 

Black holes (BHs) are implemented as `sink' particles and form in dense star-forming regions, where the gas densities are above a critical threshold $\rho_{0}$, where $\rho_{0}=1.67\times10^{-25}$g cm$^{-3}$ (equivalent to 0.1 H cm$^{-3}$), and the stellar velocity dispersion is larger than 100 kms$^{-1}$. Initial (seed) BH masses are $10^5$M$_\odot$ and BH growth occurs either via gas accretion or via mergers with other BHs. This growth is tracked self-consistently, based on a modified Bondi accretion rate \citep{Booth2009}, which is capped at Eddington. 

BHs impart feedback on ambient gas via two modes, depending on the accretion rate. For high Eddington ratios ($> 0.01$), 1.5 per cent of the energy is injected into the gas as thermal energy (a `quasar' mode). For Eddington ratios that are less than 0.01, bipolar jets are used with velocities of $10^4$ kms$^{-1}$, which constitutes a `radio' mode with an efficiency of 10 per cent \citep{Dubois2012,Dubois2014}. These parameters are chosen to produce agreement with the local M$_{BH}$-M$_{*}$ and M$_{BH}$-$\sigma_{*}$ relations \citep[e.g.][]{Haring2004}, as well as the local cosmic BH mass density \citep{Dubois2012,Volonteri2016}. 

\citet{Dubois2016} has shown that this two channel model for AGN feedback is important in influencing the morphological evolution of massive galaxies. In particular, AGN feedback is generally instrumental in stopping the persistent accretion of large amounts of gas directly from the surrounding filaments, which would otherwise result in almost all massive galaxies exhibiting large-scale discs. On the other hand, the quasar mode, which is typically triggered by events like mergers, does not completely quench the gas brought in by the merger, allowing new stars forming from this gas to maintain or enhance the rotational component of the galaxy. This is partly responsible for the formation and maintenance of massive discy systems as described below. 

Horizon-AGN reproduces an array of observational quantities that trace the evolution of stars and BHs in massive galaxies, e.g. the morphological mix of massive galaxies (M$_*$ $>$ 10$^{10.5}$ M$_\odot$) in the nearby Universe \citep{Dubois2016}, stellar mass/luminosity functions, rest-frame UV-to-near-infrared colours, the cosmic star formation history, the position of galaxies on the star formation main sequence \citep{Kaviraj2017} and the demographics of BHs over cosmic time \citep{Volonteri2016}.

In each simulation snapshot, galaxy catalogues are produced by applying the \textsc{adaptahop} structure finder \citep{Aubert2004,Tweed2009} to the star particles. Structures are identified when the local density, calculated using the nearest 20 neighbours, exceeds the average matter density by a factor of 178. A minimum of 50 particles is required for identification of structures. This imposes a minimum limit for galaxy stellar masses of $\sim2\times10^{8}$M$_{\odot}$. Merger trees are produced for each galaxy from $z=0.06$ to $z=3$, with a typical timestep spacing of $\sim130$ Myr. It is worth noting that, in our stellar mass range of interest (M$_*$ $>$ 10$^{11.4}$ M$_\odot$), merger histories of galaxies are highly complete. For example, given the mass resolution of the simulation stated above, even mergers with mass ratios down to 1:100 will be visible in the low-redshift Universe. 


\subsection{Galaxy morphology}
 
Following \citet{Martin2018b}, we define galaxy morphology using stellar kinematics, via $\sfrac{V}{\sigma}$, which is the ratio between the mean rotational velocity ($V$) and the mean velocity dispersion ($\sigma$) of the star particles in a galaxy. Objects with high values of $\sfrac{V}{\sigma}$ are rotationally-supported systems (i.e. discs), while those with lower $\sfrac{V}{\sigma}$ values are pressure-supported (spheroidal) systems. $\sfrac{V}{\sigma}$ is obtained after rotating the coordinate system, so that the $z$-axis is oriented along the stellar angular-momentum vector (calculated using all the star particles). $V$ is defined as $\bar{V_{\theta}}$, i.e. the mean tangential velocity component in cylindrical co-ordinates, while the velocity dispersion ($\sigma$) is calculated by taking the standard deviations of the radial ($\sigma_{r}$), tangential ($\sigma_{\theta}$) and vertical star particle velocities ($\sigma_{z}$) and summing them in quadrature. $\sfrac{V}{\sigma}$ is defined as:

\begin{equation}
\frac{V}{\sigma} = \frac{\sqrt{3} \bar{V_{\theta}}}{\sqrt{\sigma^2_r + \sigma^2_z + \sigma^2_\theta}}
\end{equation}

The predicted spheroid and disc fractions are compared to visual morphological classifications of massive galaxies in the low-redshift Universe \citep{Conselice2006} to calculate a threshold value of $\sfrac{V}{\sigma}$ (0.55) that separates disc galaxies from their spheroidal counterparts. In other words, galaxies with $\sfrac{V}{\sigma}>0.55$ are considered to be discs, while those with $\sfrac{V}{\sigma}<0.55$ are spheroids. In the mass range of interest in our study, the predicted morphological mix of the Universe compares well to observational values (see e.g. Figure 1 in \citet{Martin2018b}). Note that the discs we study in this paper have $\sfrac{V}{\sigma}$ values that are significantly higher than 0.55, i.e. these galaxies are firmly in the disc regime. 


\begin{center}
\begin{table*}
\centering
\begin{tabular}{|| c | c | c | c | c | c | c | c | c | c |}
\hline
\hline

1 & 2 & 3 & 4 & 5 & 6 & 7 & 8 & 9\\
\hline
Morph. & \% & $\sfrac{V}{\sigma}$ & $\log_{10}$ M$_*$ & $\log_{10}$ M$_{\textnormal{BH}}$ & log$_{10}$ acc. rt & log$_{10}$ M$_{\textrm{halo}}$ & log$_{10}$ M$_{\textrm{gas}}$ & log$_{10}$ M$_{\textrm{sf-gas}}$ \\
& & & [M$_\odot$]  & [M$_\odot$] & [M$_\odot$ yr$^{\textnormal{-1}}$] & [M$_\odot$] & [M$_\odot$] & [M$_\odot$] \\
\hline
\hline
Rejuv. discs & 7.7 & 0.68$^{0.02}$  & 11.48$^{0.02}$ & 8.90$^{0.03}$ & -2.03$^{0.04}$ & 12.79$^{0.05}$ & 10.45$^{0.06}$ & 10.27$^{0.08}$ \\
\\
Const. discs & 3.3 & 0.90$^{0.06}$  & 11.45$^{0.01}$ & 8.74$^{0.04}$ & -1.91$^{0.08}$ & 12.76$^{0.09}$ & 10.49 $^{0.04}$ & 10.39$^{0.04}$\\
\\
Spheroids & 89 & 0.15$^{0.01}$  & 11.57$^{0.01}$ & 8.98$^{0.02}$ & -2.08$^{0.04}$ & 13.19$^{0.06}$ & 10.29$^{0.03}$ & 9.86$^{0.03}$\\

\end{tabular}
 \caption{Mean properties (errors on the means are shown as superscripts) of massive galaxies. Columns: (1) morphological class, (2) percentage of a given morphological class in the massive (M$_*$ $>$ 10$^{11.4}$ M$_\odot$) galaxy population, (3) $\sfrac{V}{\sigma}$, (4) log$_{10}$ of the stellar mass, (5) log$_{10}$ of the black-hole mass, (6) log$_{10}$ of the mean black-hole accretion rate across the galaxy's lifetime, (7) log$_{10}$ of the dark-matter halo mass, (8) log$_{10}$ of the total gas mass (9) log$_{10}$ of the star-forming gas mass (i.e. gas that is dense enough to form stars, $\rho_{gas}$$>$0.1 H cm$^{−3}$).}
\label{population properties}
\end{table*}
\end{center}

\begin{center}
\begin{table*}
\centering
\begin{tabular}{|| c | c | c | c | c |}
\hline
\hline

1 & 2 & 3 & 4 & 5\\
\hline
Morph. & Redshift & Stellar mass ratio & Massive f$_{\textnormal{gas}}$ & Sat. f$_{\textnormal{gas}}$ \\
\hline
\hline
Rejuv. discs  & 0.32$^{0.07}$ & 4.29$^{0.36}$ & 0.17$^{0.02}$ & 0.33$^{0.02}$\\
\\
Const. discs & 0.36$^{0.11}$ & 4.06$^{0.59}$ & 0.19$^{0.02}$ & 0.32$^{0.03}$\\
\\
Spheroids & 0.49$^{0.02}$ & 4.44$^{0.11}$ & 0.09$^{0.01}$ & 0.23$^{0.01}$\\
\\
\end{tabular}
\caption{Mean properties (errors on the mean are shown as superscripts) of the most recent significant merger, defined as the last merger with a stellar mass ratio greater than 1:10. Columns: (1) morphological class (2) last merger redshift, (3) stellar mass ratio (4) gas fraction of the more massive galaxy, (5) gas fraction of the satellite.}
\label{merger properties}
\end{table*}
\end{center}


\subsection{Selection of extremely massive disc galaxies}

We define massive disc galaxies as those with M$_*$ $>$ 10$^{11.4}$ M$_\odot$ and $\sfrac{V}{\sigma}>0.55$ at z $=$ 0. These systems are, therefore, well beyond the knee of the galaxy stellar mass function (M$_*$ $\sim$ 10$^{10.8}$ M$_\odot$) and are in the disc regime. The total number of galaxies with M$_*$ $>$ 10$^{11.4}$ M$_\odot$ is 569 and the fraction of discs in this mass range is around 11 per cent. This fraction is in good agreement with observational work. For example, \citet{Ogle2016,Ogle2019} find that their `super-spirals' constitute $6.5\%$ ($9.2\%$ when accounting for inclination incompleteness) of galaxies with M$_*$ $>$ 10$^{11.3}$ M$_\odot$, which is consistent with our simulated values (see Table \ref{population properties}).


\subsection{Local environment}

To explore local environment we utilise information about a galaxy's host dark matter halo and its group hierarchy, i.e. whether it is a `central' or a `satellite'. Satellites are systems whose host dark matter haloes have merged into a larger halo where they currently reside. For some of our analysis we also explore the vicinity of galaxies in the cosmic web produced by Horizon-AGN, via the persistence-based filament tracing algorithm \texttt{DisPerSE} \citep{Sousbie2011}, which uses the density field estimated via a delaunay tessellation \citep{SchappetVandeWeygaert2000} of the dark matter particles. We choose a persistence of 7 sigma. \texttt{DisPerSE} identifies ridges in the density field as special lines that connect topologically robust pairs of nodes. These ridges compose the filament network of the cosmic web, and the set of all segments defining these ridges are referred to as the `skeleton' \citep{Pogosyan2009}. We refer readers to \citet{Sousbie2011} and \citet{Sousbie2011b} for more details of the \texttt{DisPerSE} algorithm and to \citet{Dubois2014} and \citet{Laigle2018} for a description of its implementation in Horizon-AGN.

\begin{figure}
\includegraphics[width=\columnwidth]{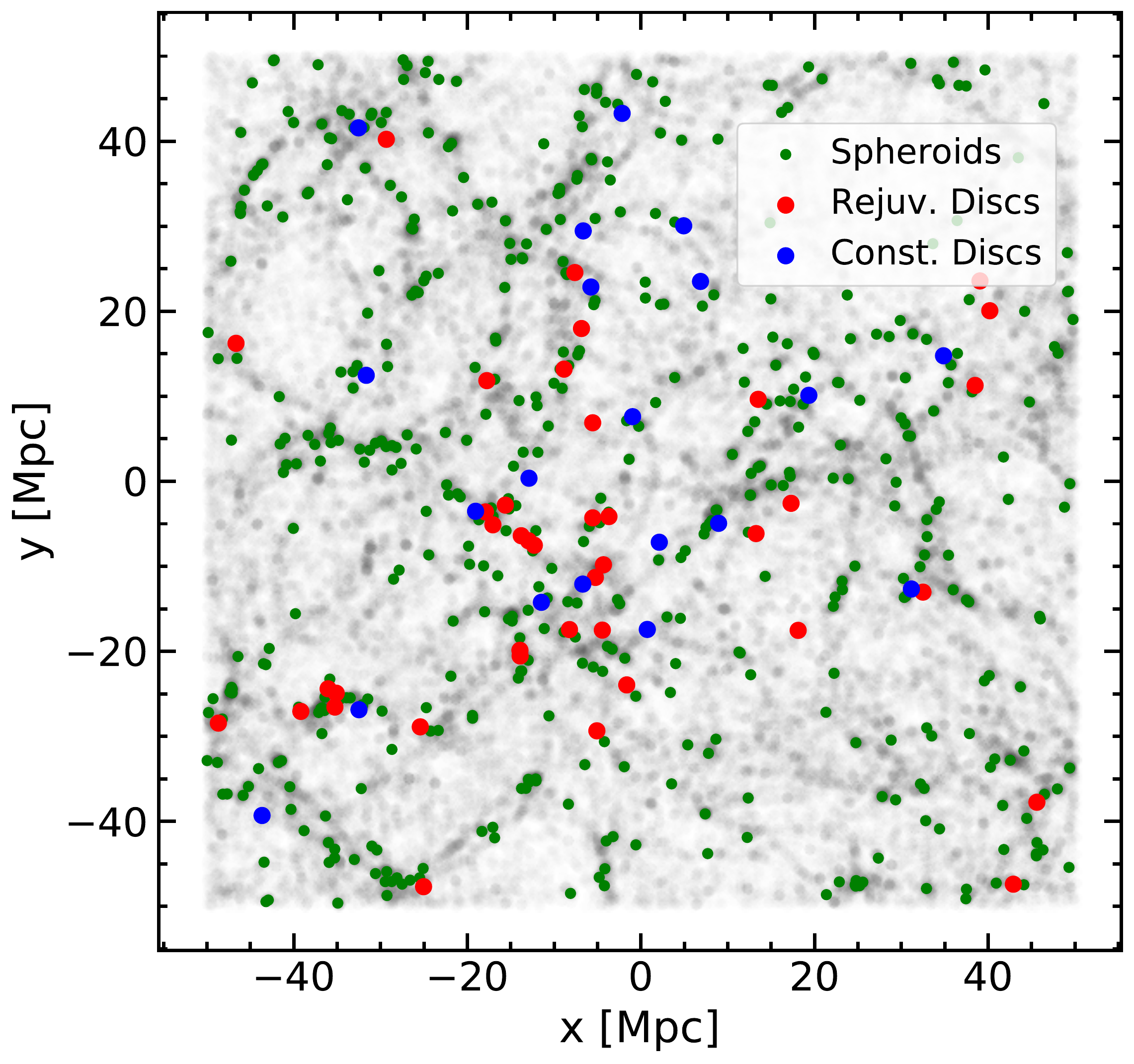}
\centering
\caption{Positions of rejuvenated discs (red), constant  discs (blue) and spheroids (green) in the cosmic web from Horizon-AGN. Grey dots indicate the general galaxy population, with darker regions indicating regions of higher density.}
\label{fig:cosmic web}
\end{figure}

We note that, out of the 64 massive discs in this study, 63 systems (i.e. more than 99 per cent) are central galaxies. This appears consistent with observational studies which indicate that many massive discs tend to be the brightest galaxies in their respective groups/clusters \citep[e.g.][]{Ledlow2001,Li20192,Bogd2018}. Figure \ref{fig:cosmic web} shows the positions of our three morphological classes (i.e. rejuvenated discs, constants discs and spheroids) in the cosmic web. Properties used to characterise local environment are tabulated in Table \ref{population properties}.

\section{How do extremely massive disc galaxies form?}
\label{sec:disc formation}

\begin{figure*}
\centering
\includegraphics[width=0.32\textwidth]{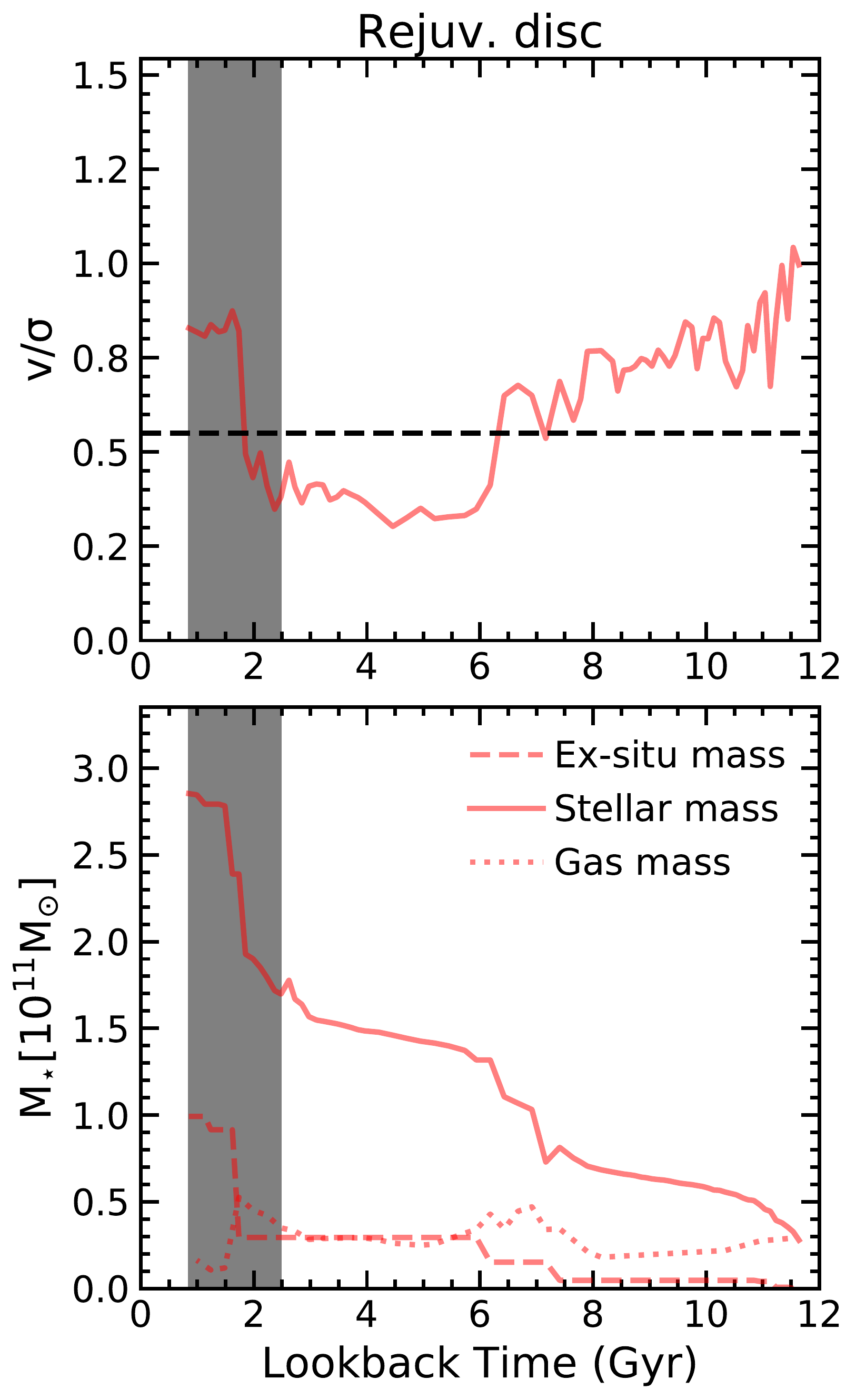}
\includegraphics[width=0.32\textwidth]{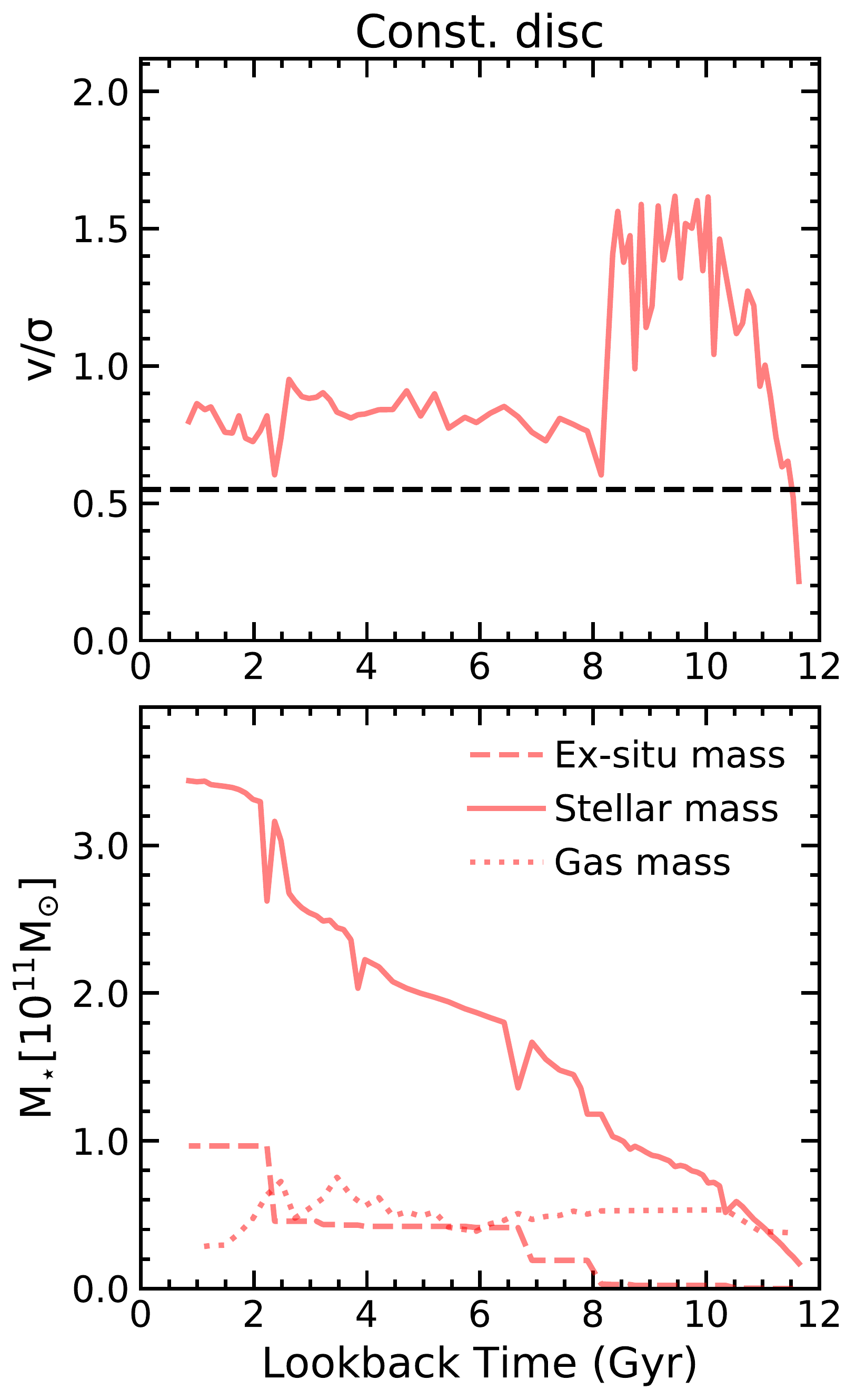}
\includegraphics[width=0.32\textwidth]{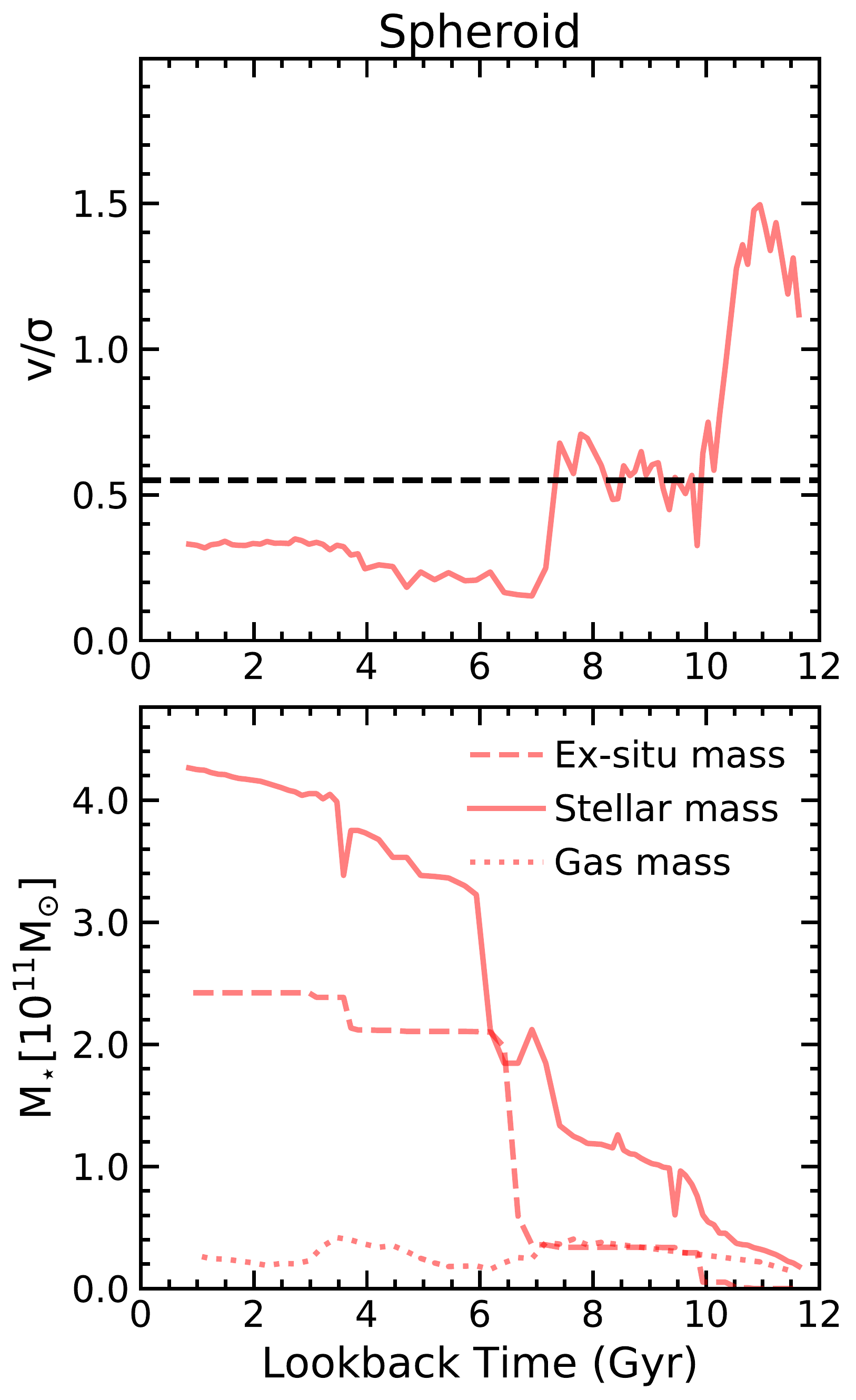}
\caption[]{The evolution of the properties of the progenitor system of massive galaxies. Each column shows the evolution of an individual galaxy. The left, centre and right-hand columns show the evolution of a rejuvenated disc, a constant disc and a massive spheroid respectively (see text in Section \ref{sec:channels} for details). The top row shows the evolution in $\sfrac{V}{\sigma}$, while the bottom row shows the change in the stellar mass (solid), ex-situ (i.e accreted) stellar mass (dashed) and gas mass (dotted) of the system. The ex-situ stellar mass shows a near-step change when mergers take place, with the magnitude of the change indicating the mass brought in by the accreted satellite. The grey region highlights the most recent merger which produces the uptick in $\sfrac{V}{\sigma}$ that moves the rejuvenated disc into the disc regime. The dotted line at $\sfrac{V}{\sigma}$ = 0.55 demarcates the spheroid and disc regimes.}
\label{fig:vsig}
\end{figure*}

\subsection{Two channels of massive disc formation}
\label{sec:channels}

We begin by illustrating, graphically, the two channels that create galaxies that are massive discs at the present day. In Figure \ref{fig:vsig} we describe the evolution in morphology and the stellar mass assembly in typical galaxies that form via these channels. The left and middle columns show the change in $\sfrac{V}{\sigma}$ (top) and the evolution of the stellar mass, ex-situ stellar mass and gas mass (bottom) for a typical system that represents each of the two channels of massive-disc formation. The right-hand column shows the same for a massive spheroid. The ex-situ stellar mass is defined as the stellar mass that is directly accreted from external objects via mergers, and not formed within the galaxy's main progenitor. 

As this figure indicates, massive discs form in one of two ways. In the first channel (left-hand column) the progenitor is initially a spheroid, until the most recent merger event causes a significant uptick in $\sfrac{V}{\sigma}$ which moves the system into the disc regime. This uptick coincides with this merger bringing in an appreciable amount of gas (since there is a coincident uptick in the gas mass) which builds a new disc component. We refer to these galaxies as `rejuvenated discs'. As we discuss below the most recent mergers that drive this rejuvenation have significant mass ratios ($>$ 1:10). It is worth noting that, in contrast, the most recent mergers in massive galaxies that exhibit spheroidal morphology (right-hand column) today are gas-poor. These mergers do not necessarily produce an uptick in $\sfrac{V}{\sigma}$ and, when they do, these upticks are not sufficient to move the system into the disc regime. Indeed $\sfrac{V}{\sigma}$ typically decreases, as is expected in mergers which are gas-poor, since the only effect of the merger is to randomise the stellar orbits and reinforce the spheroidal component of the system \citep[e.g.][]{Lotz2010,Taranu2013,Naab2014,Martin2018b}. In the second channel (middle column) the galaxy retains a disc component and remains in the disc regime throughout its lifetime. We refer to these systems as `constant discs'. 

Visual inspection of the $\sfrac{V}{\sigma}$ evolution of all massive discs indicates that the rejuvenated disc channel accounts for $\sim70\%$ of these systems (and $\sim8\%$ of all massive galaxies), with the remaining $\sim30\%$ having maintained a disc component over cosmic time (these systems comprise $\sim3\%$ of all massive galaxies). Table \ref{population properties} presents mean properties of the three morphological classes: rejuvenated discs, constant discs and spheroids. In the following sections we explore these two channels of massive disc formation in more detail. 


\subsection{The dominant channel of massive disc formation: disc rejuvenation via recent mergers}

We begin by exploring the principal channel for massive disc formation - the rejuvenation of a disc by a recent merger. We note first that the rejuvenation is always driven by mergers with significant stellar mass ratios that are greater than 1:10. Given that the change in morphology from spheroid to disc appears to be driven by the properties of the most recent significant merger, we study how the properties of these mergers differ between rejuvenated discs and their spheroidal counterparts. 

As Table \ref{merger properties} indicates, the mass ratios of the most recent significant merger, defined as the last merger a galaxy has undergone with a mass ratio greater than 1:10, are similar in both the rejuvenated discs and their spheroidal counterparts. This is not unexpected, since the merger histories of galaxies with similar stellar masses tend to be comparable, regardless of morphology \citep{Martin2018b}. The rationale for the 1:10 mass ratio threshold is that mergers below this threshold affect the system very weakly and do not produce morphological change \citep{Martin2018b}. The differences between the rejuvenated discs and spheroids are, therefore, not driven by the mass ratio of the most recent significant merger. 

However, differences arise when we consider both the gas content of this merger event and its redshift. The progenitor galaxies in mergers that create rejuvenated discs show higher gas fractions than those in their spheroidal counterparts. The median gas fractions are elevated by a factor of $\sim$2 in both the more massive progenitor and the accreted satellite. Since the mass ratios of the most recent significant mergers are similar, the absolute gas mass brought into the merger therefore tends to be higher in events that create these systems. As has been shown in previous work \citep[e.g.][]{Lotz2010,Naab2014, Lagos2018,Martin2018b}, gas-rich mergers will `spin up' merger remnants, as the gas creates a new rotationally-supported stellar component. As shown graphically in  Figure \ref{fig:vsig} (top row, left-hand column), these gas-rich recent mergers produce an uptick in $\sfrac{V}{\sigma}$, that moves the system from the spheroid to the disc regime.\footnote{For completeness, we have checked what fraction of massive spheroids which have a recent significant gas-rich (f$_{gas}>$0.3) merger remain spheroids after such an event. We find that only 2$\%$ of massive spheroids fit this description. In other words, 98$\%$ of massive spheroids that undergo such a gas-rich merger become discs.}

\begin{figure}
\centering
\includegraphics[width=\columnwidth]{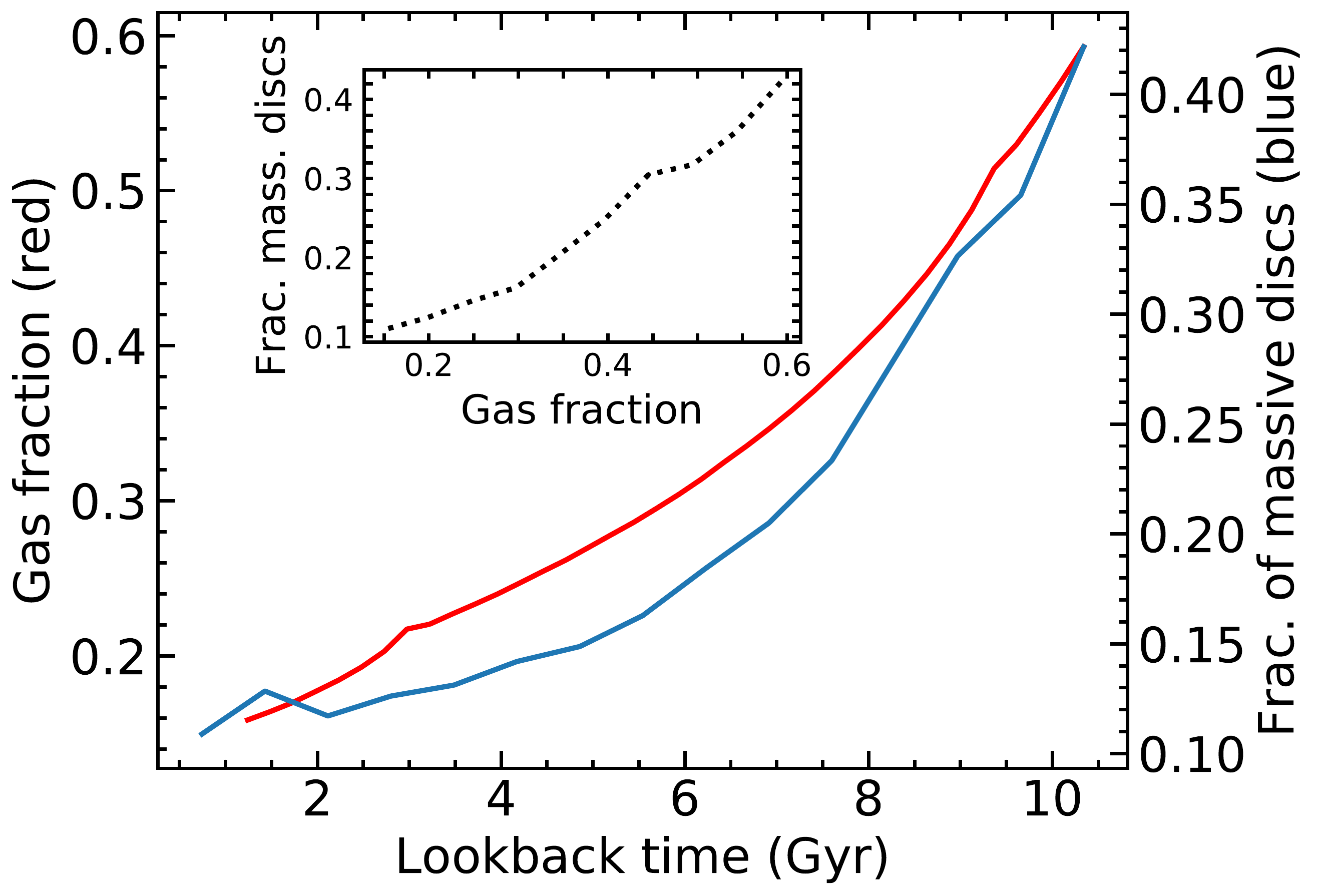}
\caption{Evolution of the gas fraction of the Universe (red) and the fraction of massive galaxies that are discs (blue), with cosmic time. The inset summarises this evolution by plotting these quantities against each other.}
\label{fig:disc fraction evolution}
\end{figure}

If rejuvenated discs, which are the dominant channel of massive-disc formation, are principally created via recent gas-rich mergers, then it stands to reason that the fraction of massive galaxies that are discs should correlate positively with the availability of gas in the Universe. Figure \ref{fig:disc fraction evolution} shows the evolution of both the gas fraction of the Universe (red) and the fraction of massive galaxies that are discs (blue) with cosmic time. At any given redshift, we define massive galaxies as those whose descendants at $z=0$ have M$_* > 10^{11.4}M_{\odot}$, with massive discs defined as massive galaxies with $\sfrac{V}{\sigma} > 0.55$ at that redshift. The inset summarises this evolution by plotting the fraction of massive galaxies that are discs against the gas fraction of the Universe. This figure confirms that, as one would expect for such a rejuvenation process, a higher gas fraction in the Universe leads to a higher fraction of massive galaxies that are discs. In other words, the frequency of massive discs, and therefore the morphological mix of galaxies at the highest stellar masses, is strongly driven by the gas fraction of the Universe.  

Analysis of the local environment (Table \ref{population properties}) indicates that rejuvenated discs typically reside in less-massive dark-matter halo masses, i.e. they inhabit less dense environments. Galaxies in these regions will be less affected by processes like ram pressure stripping and tidal heating which can remove their constituent gas \citep[e.g.][]{Vollmer2001,Johansson2009,Martin2019}. This enables these systems to merge with more gas-rich satellites which can then drive the disc rejuvenation process. 

Finally, it is worth noting that the median redshift of the last significant merger event is lower in the rejuvenated discs ($z\sim0.3$, which corresponds to a look-back time of $\sim$3.5 Gyrs) compared to that in their spheroidal counterparts ($z\sim0.49$, which corresponds to a look-back time of $\sim$5 Gyrs). This likely assists in the survival of the discy components to the present day, because less time has elapsed since the recent merger event, reducing the possibility that further significant interactions take place which could enhance the spheroidal components of these systems.

\subsection{The secondary channel of massive disc formation: disc preservation over cosmic time}

While the majority of massive discs have formed through disc rejuvenation via recent significant gas-rich mergers, a minority of this population have remained in the disc regime over cosmic time. In this section, we explore how these rare systems preserve their disc components over their lifetimes.

\begin{figure}
\centering
\includegraphics[width=\columnwidth]{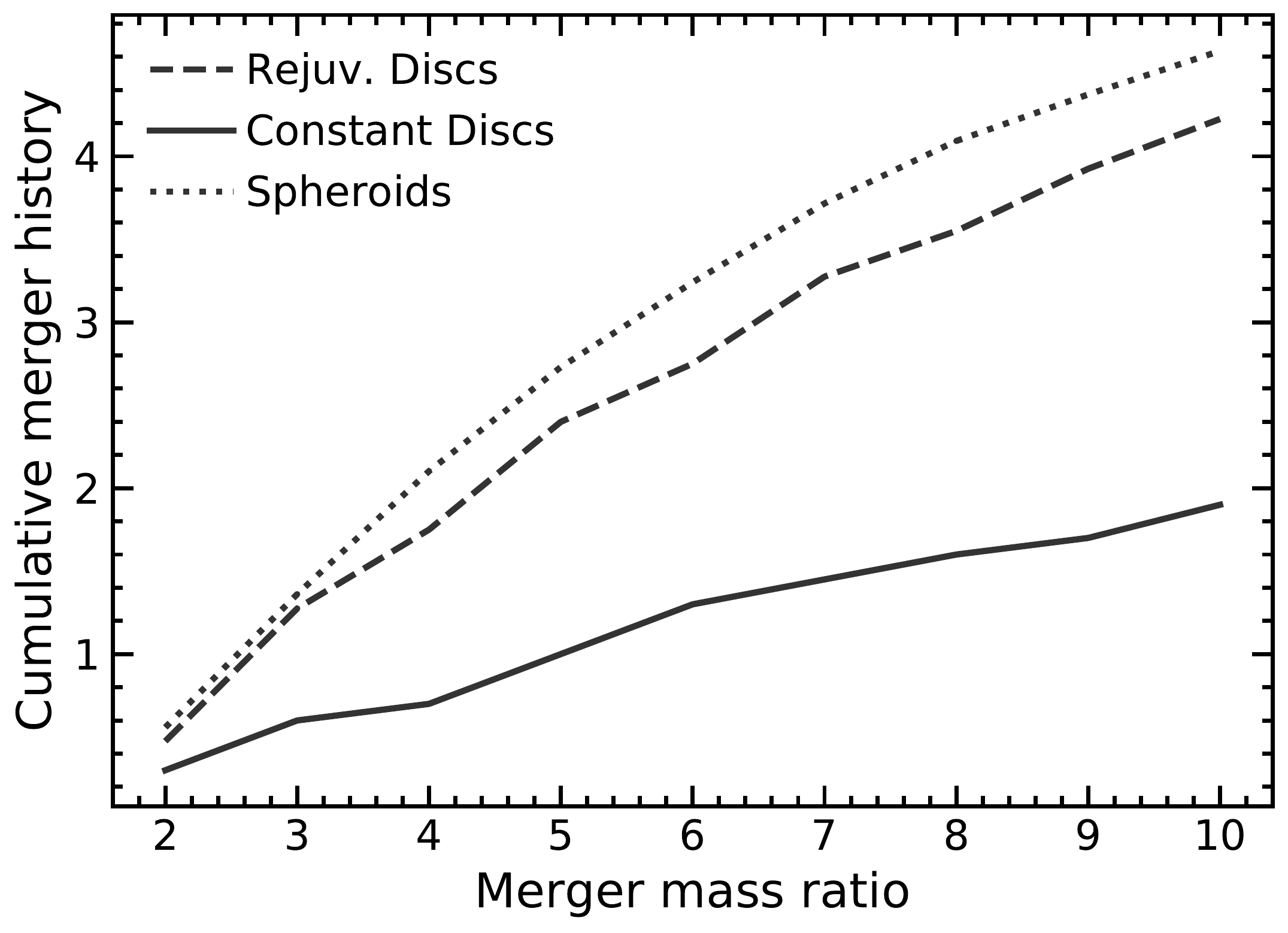}
\caption{Cumulative merger history for our three morphological classes: rejuvenated discs (dashed line), constant discs (solid line) and spheroids (dotted line). This figure presents the average number of mergers experienced by a galaxy over its lifetime, with mass ratios less than or equal to a given value, shown on the x-axis. For example, rejuvenated discs undergo, on average, $\sim$4.2 mergers with mass ratios greater than 1:10, while constant discs undergo $\sim$1.9.}
\label{fig:cumulativemerger}
\end{figure}

The most recent significant mergers in these constant discs have similar properties, e.g. merger mass ratios, the redshift of the last significant merger and the gas fractions of the merging progenitors, to those in their rejuvenated counterparts (Table \ref{merger properties}). The two sub-populations also occupy similar local environments in terms of their dark-matter halo masses. 

However, strong differences arise when we consider the cumulative merger histories of the different morphological classes across cosmic time. Figure \ref{fig:cumulativemerger} presents the average number of mergers experienced by a galaxy, over its lifetime, with mass ratios less than or equal to the value shown on the x-axis. For example, rejuvenated discs undergo, on average, 4.2 mergers with mass ratios greater than 1:10, while constant discs undergo 1.9. The rate of mergers with significant mass ratios is therefore a factor of 2.2 higher in the rejuvenated discs compared to their constant counterparts. Not unexpectedly, the rejuvenated discs have similar merger histories to their spheroidal counterparts (since they were spheroidal before their most recent significant merger). 

This anomalously quiet merger history enables the constant discs to maintain their discy components over their lifetimes. Furthermore, since mergers typically accelerate the consumption of gas \citep[e.g.][]{Martin2017,Jackson2019}, a quieter merger history also enables the system to better retain its gas, as indicated by both the higher total and star-forming gas masses in Table \ref{population properties}. 

A potential explanation for this anomalously quiet merger history is the halo bias effect \citep[e.g.][]{Borzyszkowski2017}, whereby a more massive nearby halo (typically a node in the cosmic web) effectively shields the galaxy from mergers, allowing it to continue forming stars without being disturbed by interactions. We check for the presence of a halo bias effect using the position of nodes (and other halos) in the skeleton. However, we find no evidence that this effect may be driving the properties of the constant discs. These systems are no more likely to be close to a more massive halo/node than their rejuvenated counterparts and, in most cases, their halos actually dominate the local environment. This is perhaps not unexpected because all galaxies in our sample (including the constant discs) are extremely massive. Thus, the likelihood of finding a more massive system nearby is very low. The quieter merger history of the constant discs therefore seems to be a stochastic effect, which aligns well with the extreme rarity of these systems. 


\subsection{A note about massive discy hosts of AGN}
\label{sec:AGN}

We complete our study by considering whether the massive discs studied here may provide a natural explanation for the minority of powerful AGN that appear to (surprisingly) inhabit massive disc galaxies \citep[e.g.][]{Tadhunter1992,Koff2000,Canalizo2001,Guyon2006,Madrid2006,Georgakakis2009,Morganti2011,Singh2015}.

Recall from the analysis above that the majority of massive discs are systems that are initially spheroidal, but in which discs have been rejuvenated via a recent gas-rich merger. Table \ref{population properties} indicates that the black-hole (BH) masses and accretion rates in the rejuvenated discs are predicted to be similar to those in their spheroidal counterparts, which is consistent with these systems originally being spheroids before the most recent significant merger. The BH masses in the constant discs are also comparable to the other morphological classes, although their BHs are slightly less massive (likely due to the quieter merger history) and their accretion rates are typically higher (due to the higher gas fractions in these systems). 

These theoretical predictions appear similar to what is seen in observations. For example, \citet{Tadhunter2016} show that a minority of the hosts of radio AGN at high stellar masses (M$_{*}>10^{11}$M$_{\odot}$), that have clearly discy morphologies, show the same patterns. They exhibit similarly high BH masses as their spheroidal counterparts (M$_{BH}>10^{8}$M$_{\odot}$), with broadly similar accretion rates. 

It is worth noting, however, that while in observed AGN that are hosted by massive discs, the accretion rates are slightly lower than that in their spheroidal counterparts, the opposite appears to be true in our theoretical analysis. This is largely explained by the different mass ranges considered, because our study is focused on galaxies that are more massive than those in observational studies like \citet{Tadhunter2016}. Indeed, if we reduce our stellar mass range to M$_*$ $>$ 10$^{11}$ M$_\odot$, we find that the massive discs then have lower accretion rates than their spheroidal counterparts, in line with the findings of \citet{Tadhunter2016}. Given the parallels between the massive discs in our theoretical study and their observed counterparts, the formation scenarios presented here appear to provide a natural explanation for the minority of powerful AGN that are observed to (surprisingly) inhabit disc galaxies at the highest stellar masses. 

\section{Summary}
\label{sec:summary}

Both theory and observations indicate that the morphological mix of massive galaxies changes from being disc-dominated in the early Universe to being dominated by spheroidal systems at low redshift. In the standard $\Lambda$CDM paradigm, this morphological transformation is thought to be driven by mergers. Galaxy merger histories correlate strongly with stellar mass, largely regardless of the morphology of the galaxy in question. The frequency of mergers typically increases with stellar mass, so that galaxies at the highest stellar masses tend to have the richest merger histories. However, while most massive galaxies have spheroidal morphology, a minority of systems at the highest stellar masses are, in fact, discs. Since mergers typically destroy discs and create spheroids, and the most massive galaxies typically have the richest merger histories, it is surprising that disc galaxies exist at all at the highest stellar masses (e.g. those well beyond the knee of the mass function).

We have studied the formation mechanisms of massive (M$_*$ $>$ 10$^{11.4}$ M$_\odot$) disc galaxies, in the Horizon-AGN simulation. Massive discs make up a significant minority ($\sim11\%$) of systems at such high stellar masses. We have shown that there are two channels of massive disc formation. The primary channel, which accounts for $\sim$70 per cent of these systems, is disc rejuvenation. In this channel, a massive spheroidal system experiences a recent gas-rich merger which rebuilds a disc and moves the system from the spheroid to the disc regime. The gas-rich mergers are facilitated by the fact that these systems typically inhabit less massive halos, i.e. less dense environments, than spheroidal counterparts with similar stellar masses. Galaxies in these regions are less likely to be affected by processes which deplete gas, like ram pressure and tidal stripping, making it more likely that massive galaxies can have gas-rich mergers.  

In the secondary channel, a massive disc remains in the disc regime over its entire lifetime. The maintenance of the disc is the result of an anomalously quiet merger history, whereby these systems undergo a factor of $\sim$2 fewer mergers with mass ratios greater than 1:10 than other galaxies with similar stellar masses. Since mergers accelerate gas consumption, a quieter merger history also enables the galaxy to retain its gas reservoir more easily, further enabling it to maintain its disc component over its lifetime. The dominance of the rejuvenation channel means that the fraction of massive galaxies that are discs is progressively larger at higher redshift, since the Universe is more gas-rich. The morphological mix of galaxies at the very highest stellar masses (at any epoch) is therefore a strong function of the gas fraction of the Universe. 

Finally, we have shown that the BH masses and accretion rates of massive discs are similar to those in their spheroidal counterparts. The formation mechanisms described here therefore provide a natural explanation for the minority of powerful AGN that are (surprisingly) found in disc galaxies.


\section*{Acknowledgements}
We are grateful to Stas Shabala and Lorenzo Posti for many interesting discussions. RAJ acknowledges support from the STFC [ST/R504786/1]. GM acknowledges support from the STFC [ST/N504105/1]. SK acknowledges a Senior Research Fellowship from Worcester College Oxford. CL is supported by a Beecroft Fellowship. JD acknowledges funding support from Adrian Beecroft, the Oxford Martin School and the STFC. This research has used the DiRAC facility, jointly funded by the STFC and the Large Facilities Capital Fund of BIS, and has been partially supported by grant Spin(e) ANR-13-BS05-0005 of the French ANR. This work was granted access to the HPC resources of CINES under the allocations 2013047012, 2014047012 and 2015047012 made by GENCI. This work is part of the Horizon-UK project.



\bibliographystyle{mnras}
\bibliography{bib} 






\bsp	
\label{lastpage}
\end{document}